\documentclass[12pt]{iopart}
\usepackage[utf8x]{inputenc}
\usepackage{epsfig}
\expandafter\let\csname equation*\endcsname\relax
\expandafter\let\csname endequation*\endcsname\relax
\usepackage{amsmath}
\usepackage{color}
\usepackage{verbatim}

\newcommand{\refB}[2]{\color{green}#2\textsuperscript{(#1)} \color{black}}
\newcommand{\refe}[1]{\color{red}#1 \color{black}}
\newcommand{\apgt}{\ {\raise-.5ex\hbox{$\buildrel>\over\sim$}}\ }
\newcommand{\aplt}{\ {\raise-.5ex\hbox{$\buildrel<\over\sim$}}\ }
\newcommand{\inod}{\tilde \omega_{f}}
\newcommand{\inad}{\tilde A_{f}}
\newcommand{\ino}{\omega_{f}}
\newcommand{\ina}{A_{f}}
\newcommand{\lfoa}{A_{0}}
\newcommand{\lfoo}{\omega_{0}}
\newcommand{\inom}{\omega_{f_m}}
\newcommand{\inam}{\hat A_{f_m}}

\newcommand{\Sm}{S_m}
\newcommand{\lfoom}{\omega_{0_m}}
\newcommand{\lfoam}{A_{0_m}}

\usepackage{ulem}

\newif\ifscaling
\scalingfalse

\begin{document}

\title{Low-frequency oscillations in narrow vibrated granular systems.}
\author{N Rivas, S Luding and A R Thornton}
\address{Multi Scale Mechanics (MSM), MESA+, CTW, University of Twente, PO Box 217, 7500 AE Enschede,
The Netherlands.}
\ead{N.A.Rivas@utwente.nl}

\begin{abstract}
We present simulations and a theoretical treatment of vertically vibrated granular media. The systems considered are confined in narrow quasi-two-dimensional and quasi-one-dimensional (column) geometries, where the vertical extension of the container is much larger than both horizontal lengths. The additional geometric constraint present in the column setup frustrates the convection state that is normally observed in wider geometries. This makes it possible to study collective oscillations of the grains with a characteristic frequency that is much lower than the frequency of energy injection. The frequency and amplitude of these oscillations are studied as a function of the energy input parameters and the size of the container. We observe that, in the quasi-two-dimensional setup, low-frequency oscillations are present even in the convective regime. This suggests that they may play a significant role in the transition from a density inverted state to convection. Two models are also presented; the first one, based on Cauchy's equations, is able to predict with high accuracy the frequency of the particles' collective motion. This first principles model requires a single input parameter, i.e, the centre of mass of the system. The model shows that a sufficient condition for the existence of the low-frequency mode is an inverted density profile with distinct low and high density regions, a condition that may apply to other systems too. The second, simpler model just assumes an harmonic oscillator like behaviour and, using thermodynamic arguments, is also able to reproduce the observed frequencies with high accuracy.
\end{abstract}


\maketitle

\section{Introduction} 

Vibrated beds of granular materials present a wide range of different behaviours: phase separation \cite{olafsen_clustering_1998,melby_dynamics_2005}, Faraday-like pattern formation instabilities \cite{melo_transition_1994,luding_simulations_1996}, heap formation and convection \cite{tennakoon_vertical_1998,medved_connections_2002}, segregation \cite{ahmad_observation_1973,kudrolli_size_2004}, clustering \cite{luding_cluster-growth_1999} and periodic cluster expansions \cite{rivas_sudden_2011}, as well as many others. These systems generally present a remarkable collection of distinct nonequilibrium inhomogeneous stable states for relatively small changes in the energy injection parameters. Hence, they are specially suited for the study of nonequilibrium phase transitions, as well as non-linear phenomena in general. Careful analysis of the microscopic mechanics behind the different transitions improves the comprehension of the complex dynamics present in driven granular systems. This gives further insight into when, and until what point, granular media behave like classical gases, fluids or solids, or whether they require an altogether different theoretical approach.

As can be seen by the aforementioned studies, the geometry of the system plays a fundamental role in determining the phenomena. Just by reducing the effective dimensionality of the system it becomes possible to observe behaviour not easily identifiable in fully three dimensional systems. The natural approach of study is then confining the grains to quasi-two-dimensional systems, where also particle-tracking methods become possible. Our study is inspired by one specific quasi-two-dimensional geometry that presents several distinct states in the energy injection parameter space: a vertical narrow box. That is, we focus on a vertically vibrated Hele-Shaw cell with the walls parallel to gravity, inside which the grains are located. The first reported classification of the different states present in this geometry was realised by Thomas et al. \cite{thomas_identifying_1989}, in what would now be considered the low energy injection limit. Research then focused on the wave-like dynamics of the granular bed, and its variations with the frequency and the amplitude of oscillation \cite{douady_subharmonic_1989, wassgren_vertical_1996}. It was with simulations that the energy input was considerably increased, and the existence of a density inverted state was first reported \cite{meerson_close-packed_2003}. This state, named Leidenfrost after the analogous water-over-vapour phenomena \cite{leidenfrost_aquae_1756}, was then experimentally studied in depth by Eshuis et al. \cite{eshuis_granular_2005, eshuis_phase_2007}, as well as the buoyancy driven convection regime that is observed for higher energy inputs.

\begin{figure}
  \begin{center}
  \includegraphics[scale=0.53]{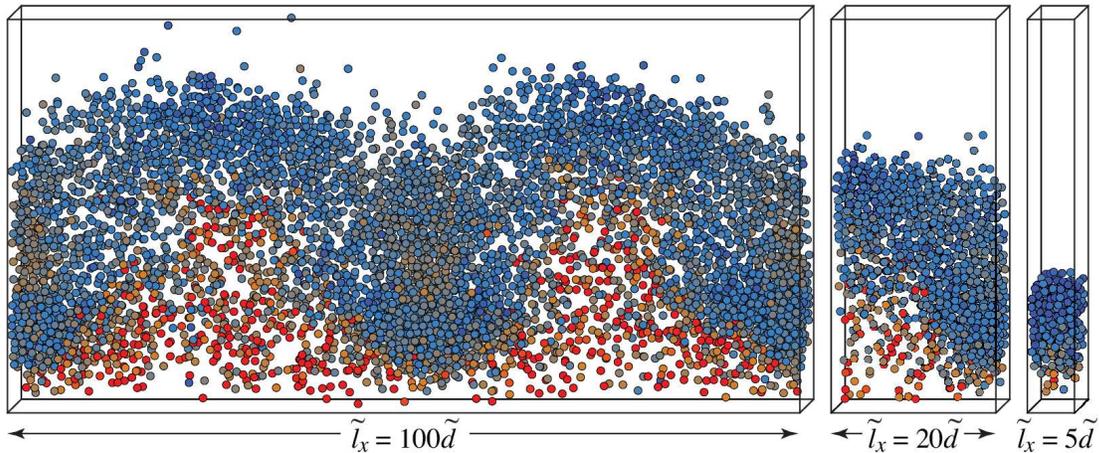}
  \end{center}
  \caption{Snapshots of three vertical narrow box systems with the same number of filling layers $F = N\tilde d^2/\tilde l_x \tilde l_y = 12$, with N the total number of particles, $\tilde d$ the (dimensional) diameter of the spherical particles, and $\tilde l_x$ and $\tilde l_y = 5\tilde d$ the (dimensional) width and depth of the container, respectively; and energy injection parameters, but different widths $\tilde l_x$. From left to right, $\tilde l_x = 100\tilde d$, $\tilde l_x = 20\tilde d $, and $\tilde l_x = 5\tilde d$. The rightmost corresponds to the \textit{column} geometry. Particles are coloured according to their kinetic energy.}
 \label{figSetup}
\end{figure}

In the following simulational study the dimensionality of the vertical, narrow box is progressively reduced until both the width and depth are only five particle diameters wide, rendering the system effectively one-dimensional (see Figure~\ref{figSetup}). We refer to this last setup as the granular column. The first direct consequence of this further confinement is the frustration of the horizontally inhomogeneous states present in the wider geometries. Particularly, the suppression of convection makes it possible to directly observe the grains collectively oscillating at a much lower frequency than the energy injection frequency. Appropriately, we call these oscillations \textit{low-frequency oscillations} (LFOs). Effective frequencies and amplitudes are defined and studied in the container length and energy injection parameter space. We then argue that LFOs are an essential feature of the dynamics of the narrow vibrated geometry, but it is only in the quasi-one-dimensional column setup that they can be easily isolated from the other collective grain movement of convection. Simulational measurements confirm this, as well as a continuum description of the system, which captures the correct frequency response for high energy inputs. The frequency behaviour is actually analogous to a forced harmonic oscillator, and is obtained mainly by considering a vibrated media with a high density region suspended over a low density one. This density inverted situation is indeed present, to different extents, in both the Leidenfrost and the convective regimes.

\section{Simulations} 

Simulations are performed using an event-driven molecular dynamics algorithm \cite{lubachevsky_how_1991}. In this approach, particles move freely under the effect of gravity until an event take place, namely, a collision with another particle or a wall. The motion of the particles in between successive events does not have to be simulated: if their trajectory equation is known, time can be advanced in variable steps. This makes event-driven simulations considerably faster than usual soft-particle simulations, where time is advanced at constant steps, independent of particle interactions. However, the need of having an analytical expression for the particle motion is a strong condition that limits the possible interaction between particles. In the following, we consider the most common approach: perfect hard-spheres, which imply binary collisions and no overlap or long-range forces between particles. Collisions are then modelled by normal and tangential restitution coefficients, $\epsilon_n = \epsilon_t = 0.95$, and also static and dynamic friction coefficients, $\mu_s = \mu_d = 0.1$ \cite{luding_granular_1995}. In order to avoid inelastic collapse we use the TC model~\cite{luding_how_1998}, with a constant value $t_c = 10^{-6} (\tilde d/\tilde g)^{1/2}$, with $\tilde d$ the (dimensional) diameter of the spheres, and $\tilde g$ the (dimensional) gravity. (In the following, quantities without a tilde are dimensionless). That is, collisions between two entities are considered elastic if they occur more frequent than $10^{-6}$ gravity timescale units. These values are known to reproduce complex behaviour observed in similar vibrated setups using stainless-steel spheres of $\tilde d = 1$~mm to $\tilde d = 5$ mm in diameter \cite{rivas_sudden_2011,clerc_liquid-solid-like_2008}.

The setup consists of an infinitely tall container of width $\tilde l_x$ and depth $\tilde l_y$ inside which the grains can move. The boundaries of the container are considered solid, and have the same collision parameters as particle-particle collisions. The whole box (both the bottom and the side walls) is vertically vibrated in a bi-parabolic, quasi-sinusoidal way with a given frequency $\inod$ and amplitude $\inad$. The use of a quadratic instead of a sine function gives a considerable speed advantage in simulations, as the prediction of collision times with the moving walls becomes substantially faster. Test simulations were performed using a sine function for exemplary cases, and no significant differences were observed \cite{mcnamara_energy_1998}. Furthermore, we considerably varied the collision parameters and found essential phenomena to be robust, although friction was observed to play a significant role in the dynamics.

We now introduce dimensionless variables, which will be used for the rest of the text. The depth of the box is fixed, $l_y \equiv \tilde l_y/ \tilde d = 5$, and the horizontal width $l_x \equiv \tilde l_x/ \tilde d$ is varied in the $[5,100]$ region. $N$ is always taken so that the number of filling layers $F \equiv N \tilde d^2 / \tilde l_x \tilde l_y = N / l_x l_y = 12$, which implies that $N$ varies in the $[300,6000]$ range. Three different oscillation amplitudes are considered, $\ina \equiv \inad / \tilde d \in \{0.4,1.0,4.0\}$. This allows us to compare with previous results, obtained for $\ina = 4.0$, as also to extrapolate to lower amplitudes, where the vibrating bottom wall can be considered as a spatially fixed source of energy (i.e. a temperature boundary condition).

The dimensionless gravity timescale is given by $t_g \equiv \tilde t / \tilde t_g$, with $\tilde t_g \equiv (\tilde d/\tilde g)^{1/2}$. Correspondingly, the dimensionless oscillation frequency $\ino$ is scaled as $\ino \equiv \inod \tilde t_g = \inod (\tilde d/\tilde g)^{1/2}$. Nevertheless, it is almost always more meaningful to measure time in periods of box oscillations, $\tilde T = 2\pi/\inod$, and thus we use $t \equiv \tilde t / \tilde T$. In order to compare simulations with different energy injection parameters the dimensionless shaking strength is used, $S \equiv \inad^2 \inod^2 /\tilde g \tilde d = \ina^2 \ino^2$. Finally, the mass scale $m = \tilde m / \tilde m_p$ is set to unity by taking $\tilde m$ as the mass of one particle, $\tilde m = \tilde m_p$.

Simulations are generally run for $10^5\,T = 10^5 (2 \pi / \ino)$, unless otherwise stated. Particles are initially arranged in a  strongly perturbed, low density crystalline state. We confirmed that this initial configuration has no influence on the steady dynamics by running a few simulations using the end state of the previous simulation as the initial configuration. Contrary to the experiments realised in \cite{eshuis_phase_2007}, where the frequency of shaking is continuously increased, the energy injection parameters are kept fixed during any given simulation. 

\subsection{Phase Space}

\begin{figure}
 \begin{center}
  \includegraphics[width=0.49\columnwidth]{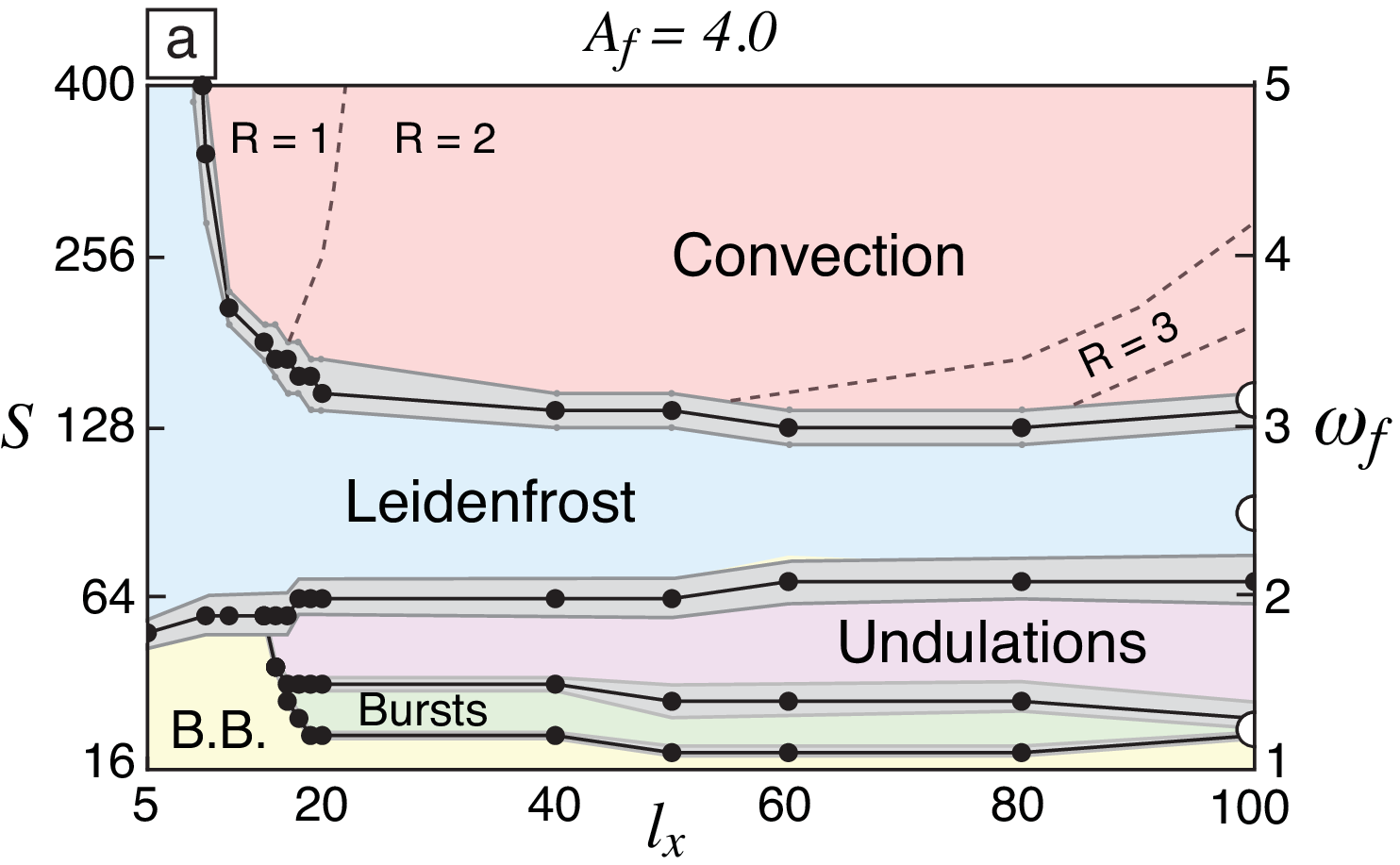}
  \includegraphics[width=0.49\columnwidth]{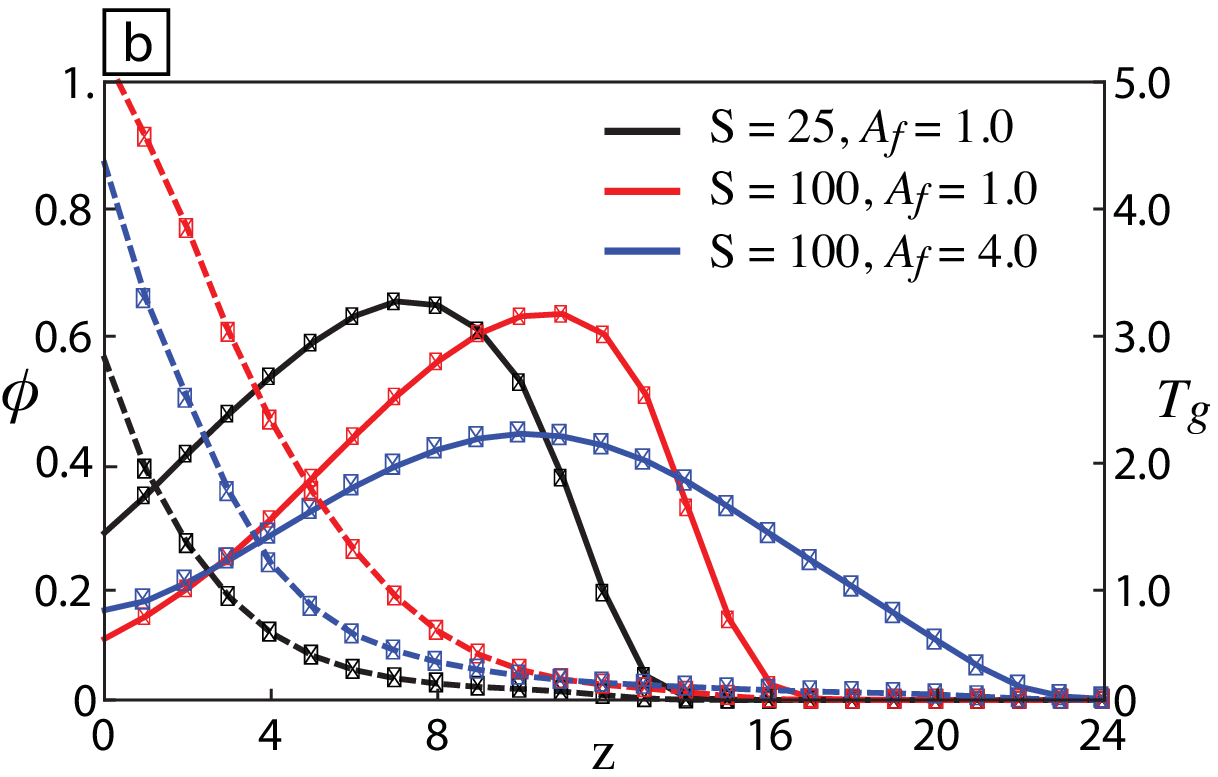}
  \end{center}
 \caption{	(a)
 			Phase diagram of the vertically vibrated system in the dimensionless container width ($l_x$) and shaking strength ($S$) space, for fixed box oscillation amplitude $\ina = 4.0$. The equivalent box oscillation frequency $(\ino)$ is shown on the right axis. All previously reported states are seen: bouncing bed (b.b, yellow), bursts (green), undulations (purple), Leidenfrost (blue) and convection (red). Transition regions are shown in gray, and are defined by the regions of bistability of every pair of states. The borders between different numbers of convective rolls ($R=1,2,3$ and $4$) is also delimited (dashed lines).
	       (b)
	      	Packing fraction $\phi$ (solid) and granular temperature $T_g$ (dashed) vertical profiles, for the exemplary amplitudes and frequencies shown in the inset, and $l_x = 50$. All these systems are in the Leidenfrost state.}
 \label{figPhase1}
\end{figure}

In order to validate our simulations, and explore further previous research, we first focus on the $\ina = 4.0$ case, where the comparison with previous experiments and soft-particle simulations undertaken by Eshuis et al. \cite{eshuis_phase_2007} is straightforward. Event-driven simulations are able to reproduce all previously observed states, as can be seen in the phase diagram in the $\{l_x,S\}$ space presented in Figure \ref{figPhase1}a. Furthermore, a quantitative comparison is possible by looking at the transition points in the $l_x = 100$ case, where the experiments were realised. There is excellent quantitative agreement, within $5$\%, for the bouncing bed-undulations and the Leidenfrost-convection transition points, but a $30$\% error in the undulations-Leidenfrost one. The deviations could in part be explained by the nature of the transitions, as they are not sharp and are seen to present wide ranges of metastability. This makes it harder to define a precise transition point value, and motivates the use of transition regions, which we show in gray.

As can be seen from Figure \ref{figPhase1}a, for $l_x > 20$, and as $S$ is increased (keeping $\ina = 4.0$), the system goes through a sequence of different non-equilibrium stable states: bouncing bed, bursts, undulations, Leidenfrost, convection and gaseous ($S > 400$, not shown). Some of these states disappear or appear as $l_x$ and $\ina$ are varied, but their relative order remains. In the following, we will focus on the Leidenfrost and convective states, where LFOs take place. For a description of the other states, we refer the reader to~\cite{thomas_identifying_1989,eshuis_phase_2007}.


The density inverted \textit{Leidenfrost} state owes its name to the analogous liquid-over-vapour phenomena, where a thin layer of vapour over a hot surface significantly slows the evaporation of the droplet above it, by keeping it floating over the hot surface \cite{leidenfrost_fixation_1966}. Figure \ref{figPhase1}b shows the packing fraction $\phi$ and the granular temperature $T_g$ as a function of $z$, for different amplitudes and frequencies of oscillation, all in the Leidenfrost state. The granular temperature is defined as twice the fluctuating kinetic energy per degree of freedom: $T_g = \tfrac{2}{3}\sum_i(\vec v_i - \vec V(r_i))^2$, where $r_i$ is the position of particle $i$, $v_i$ its velocity, and $V$ the average velocity field. Indeed, a low temperature, high density region is suspended over a low density, high temperature one. Notice that the difference in density between the solid and gaseous regions is greater for higher $\ino$ (red vs. black), but lower for higher $\ina$ (blue vs. red): these features will be relevant in our model discussion for the validity regions of a density profile approximation.

When $S$ is further increased, the density of the solid region is seen to progressively decrease, leading to a buoyancy driven convective state (see Figure \ref{figSetup}). Horizontal homogeneity is lost, leading to low density regions where particles go up and circulate around high density regions, where particles agglomerate and move mainly in the horizontal directions, towards the low density regions. The number of convection rolls ($R$) diminishes with increasing $\ino$, until the energy input is so high that particle motion is essentially uncorrelated and the system enters the gaseous state ($S > 400$, data not shown).

\begin{figure}
 \begin{center}
  \includegraphics[width=0.59\columnwidth]{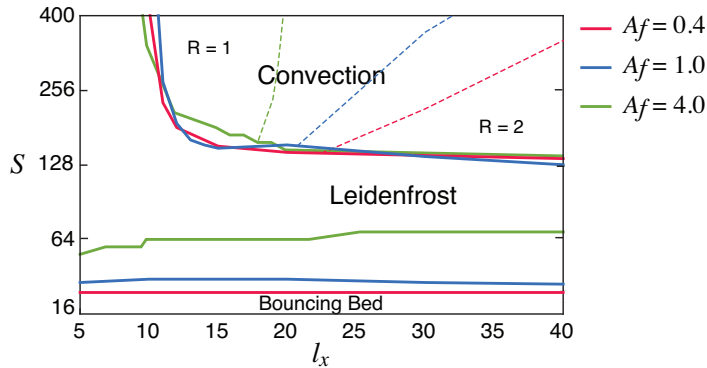}
  \end{center}
 \caption{
 	Phase diagram of the vertically vibrated narrow box in the shaking strength ($S$) and container width ($l_x$) space, for different oscillation amplitudes, as shown in the legend.
 }
 \label{figPhase2}
\end{figure}

We now turn our attention to the lower amplitude regions. Figure \ref{figPhase2} shows a phase diagram again in the $\{l_x,S\}$ parameter space, for different shaking amplitudes $\ina$. As observed previously \cite{eshuis_phase_2007}, and confirmed here for a wider range of parameters, the dimensionless shaking strength $S$ is a better parameter than the dimensionless acceleration, $\Gamma \equiv \tilde \ina \tilde \ino^2 / \tilde g = A_f \omega_f^2$ for the characterisation of the Leidenfrost-convection transition. On the other hand, the transition points of bouncing bed-Leidenfrost (or undulations-Leidenfrost for $\ina = 4.0$) vary significantly with $S$, but stay within $5\%$ when compared in $\Gamma$. In general terms, the most significant influence of reducing $\ina$ is the disappearance of the bursts and undulations states; the large amplitude of the box oscillation plays a dominant role in the dynamics of these states.

We briefly remark that simulations were done until $l_x = 400$, and no new states were observed, except for the coexistence of convection and Leidenfrost states for $l_x \ge 200$. The possibility of this coexistence provides new insight into the nature of the Leidenfrost-convection transition, and suggests further research.

If the length is reduced further, below the $l_x = 20$ limit, the frequency needed to trigger convection progressively increases, until at $l_x \sim 10$ (a value slightly dependent on $\ina$, see Figure \ref{figPhase2}) no convection was observed even for $S = 10^4$. For $\ina = 4.0$, undulations and bursts are also frustrated by the small size of the container. It is in this geometry that it becomes possible to observe the Leidenfrost state for higher $S$, where low-frequency oscillations (LFOs) can be directly observed and eventually, as $S$ is increased, dominate the collective dynamics of the system.

\subsection{Low-Frequency Oscillations (LFOs)}

\begin{figure}
 \begin{center}
 \includegraphics[width=0.49\columnwidth]{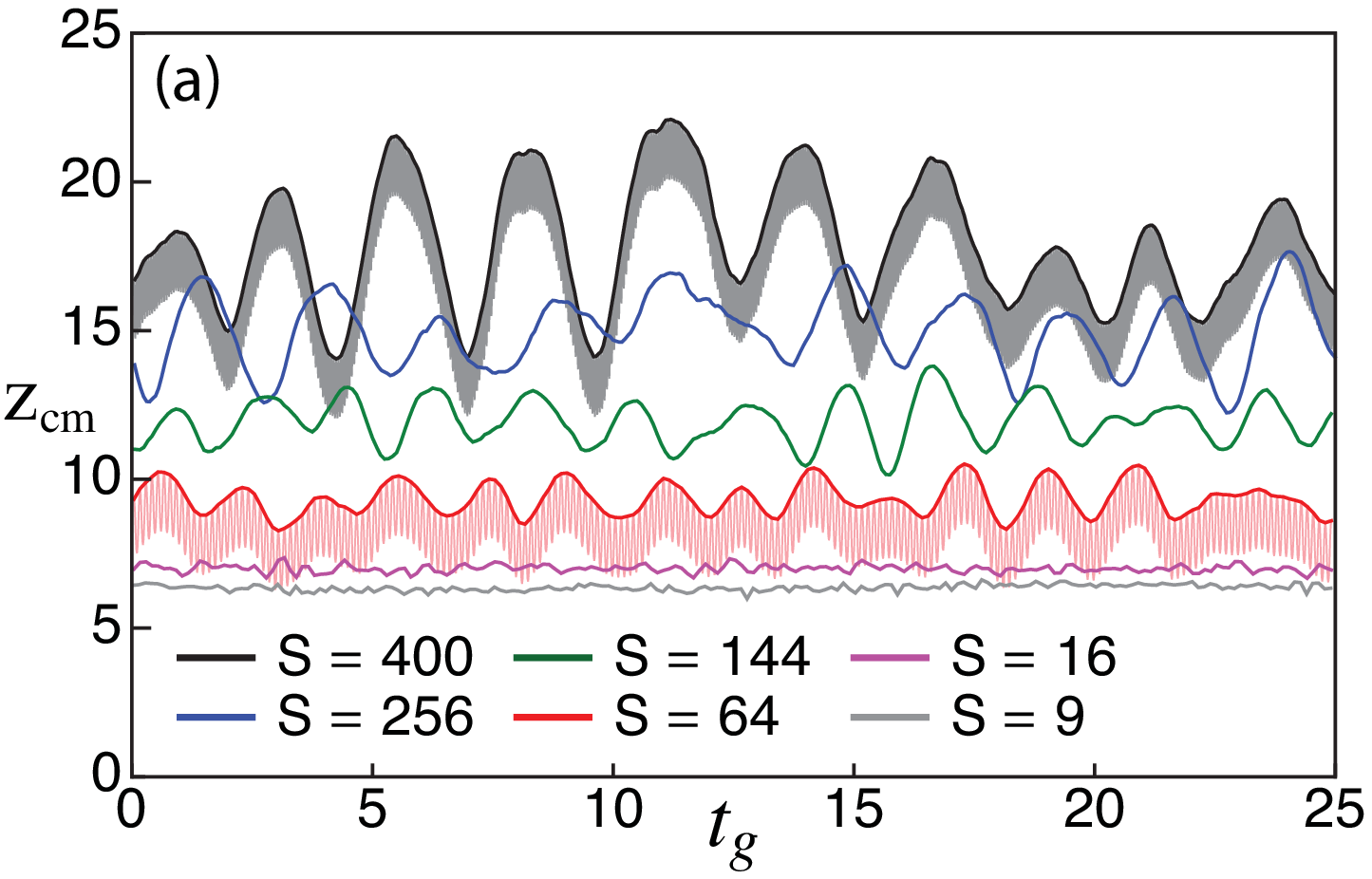}
 \includegraphics[width=0.49\columnwidth]{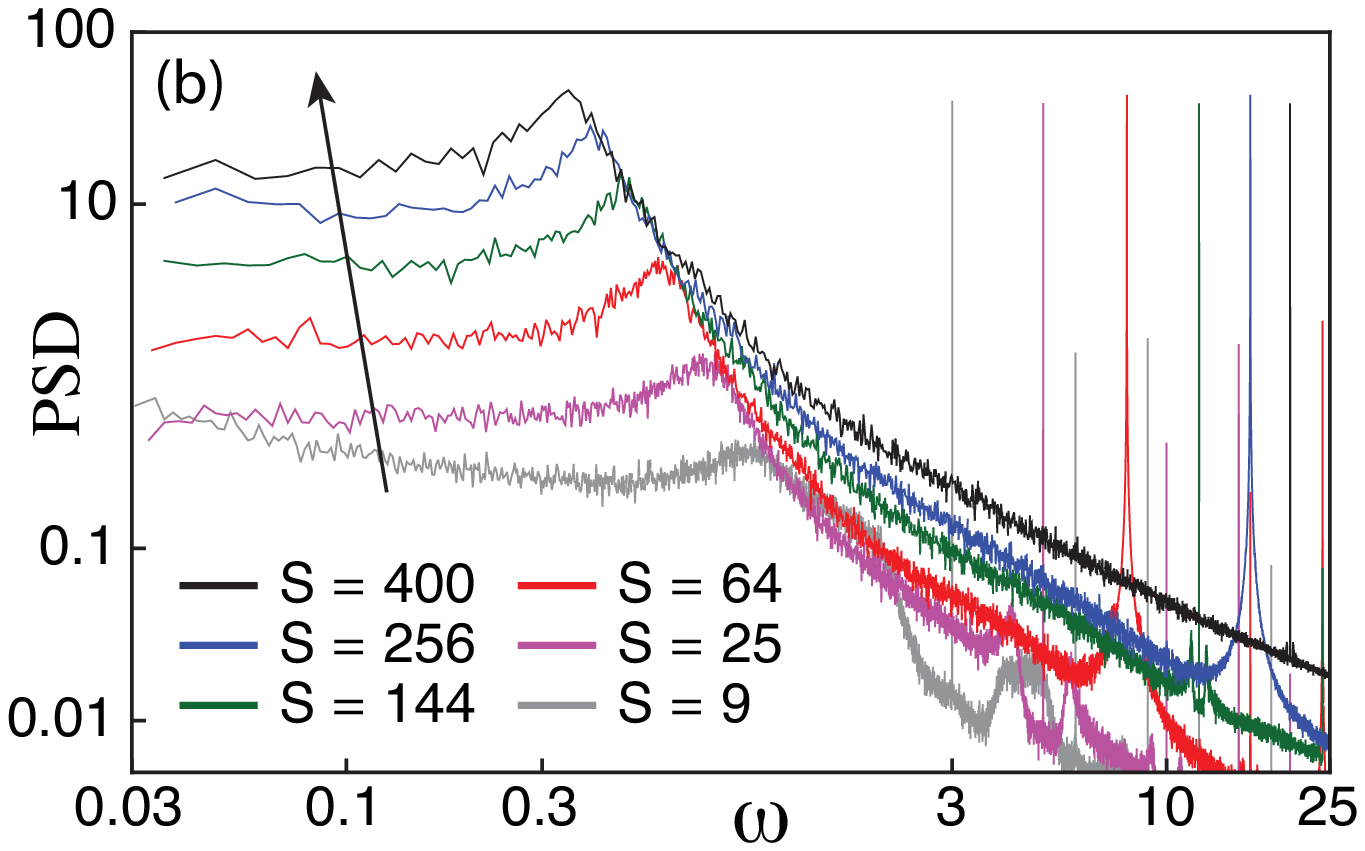}
 \end{center}
 \caption{	(a)
 			Centre of mass evolution, $z_{cm}(t)$, for $\ina = 1.0$ and different dimensionless shaking strengths $S = \ino^2$, as a function of time in gravity timescale units $t_g = \tilde t\,(\tilde g/\tilde d)^{1/2}$. The light colour data are taken with sub-period resolution, while dark colour data are taken every oscillation cycle at the point of maximum wall amplitude.
 		(b)
 			Fast Fourier transform of the centre of mass of the particles, $z_{cm}(t)$, for $\ina = 1.0$ and several different $S$. The arrow indicates the direction of increasing $S$. Different amplitudes, not shown, present the same qualitative behaviour.}
 \label{figZCM}
\end{figure}

Finally, we reach the column limit, where $l_x = l_y = 5$. In order to study LFOs the evolution of the vertical centre of mass of the particles is considered, $z_{cm}(t)$. Figure \ref{figZCM}a shows $z_{cm}(t)$ for fixed $\ina = 1.0$ and several different $S$. For comparison, non-stroboscopical and stroboscopical $z_{cm}(t)$ are shown for the $S = 64$ and $S = 400$ cases: the distinct high and low frequencies become immediately recognisable. The amplitude of the oscillations is seen to increase from the $\Delta z_{cm} \approx 1$ to the $\Delta z_{cm} \approx 10$ order, and present an appreciable regularity in time. While at $S = 64$ both oscillations are comparable in amplitude, and thus very hard to identify from direct observation, at $S = 400$ they have become clearly differentiable. Although LFOs are seen to be fairly chaotic (recall that there are only 300 particles in the column geometry, hence fluctuations play a leading role), we characterise them by a constant amplitude $\lfoa$ and a single frequency $\lfoo$, as an initial first order description.

First, let us focus on the frequency of the LFOs, $\lfoo$, which is clearly recognisable from the power spectra of $z_{cm}(t)$, presented in Figure \ref{figZCM}b. The spectra are obtained by taking the discrete fast Fourier transform of $z_{cm}(t)$ over $20000T$ after an initial transient of $1000T$, with a sampling rate of $0.05T$. An average is then taken over $10$ simulations with identical parameters but different initial conditions; although the shape and peaks are already recognisable from single simulations, the ensemble averaging reduces the noise considerably. The time window, the sampling rate and the transient time were varied and no significant differences were observed.

\begin{figure}
 \begin{center}
  \includegraphics[width=0.32\columnwidth]{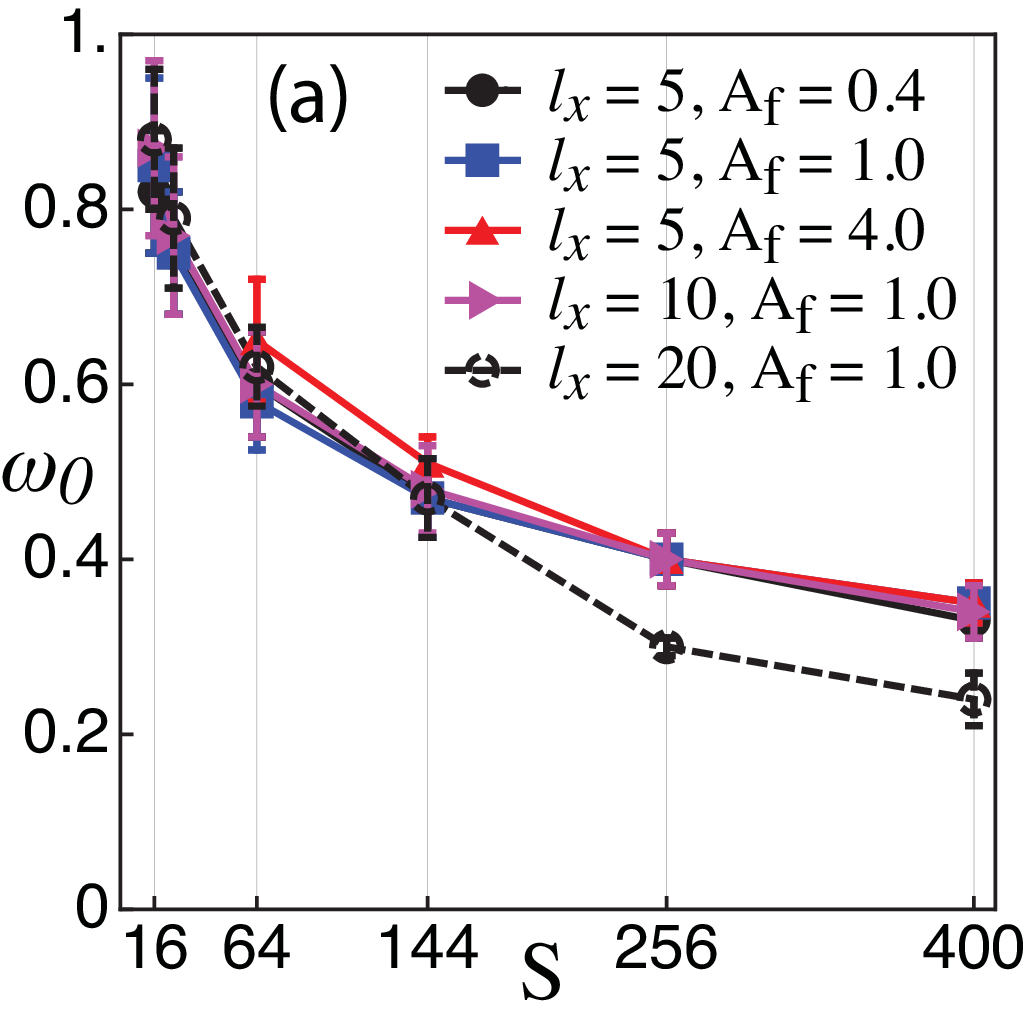}
  \includegraphics[width=0.32\columnwidth]{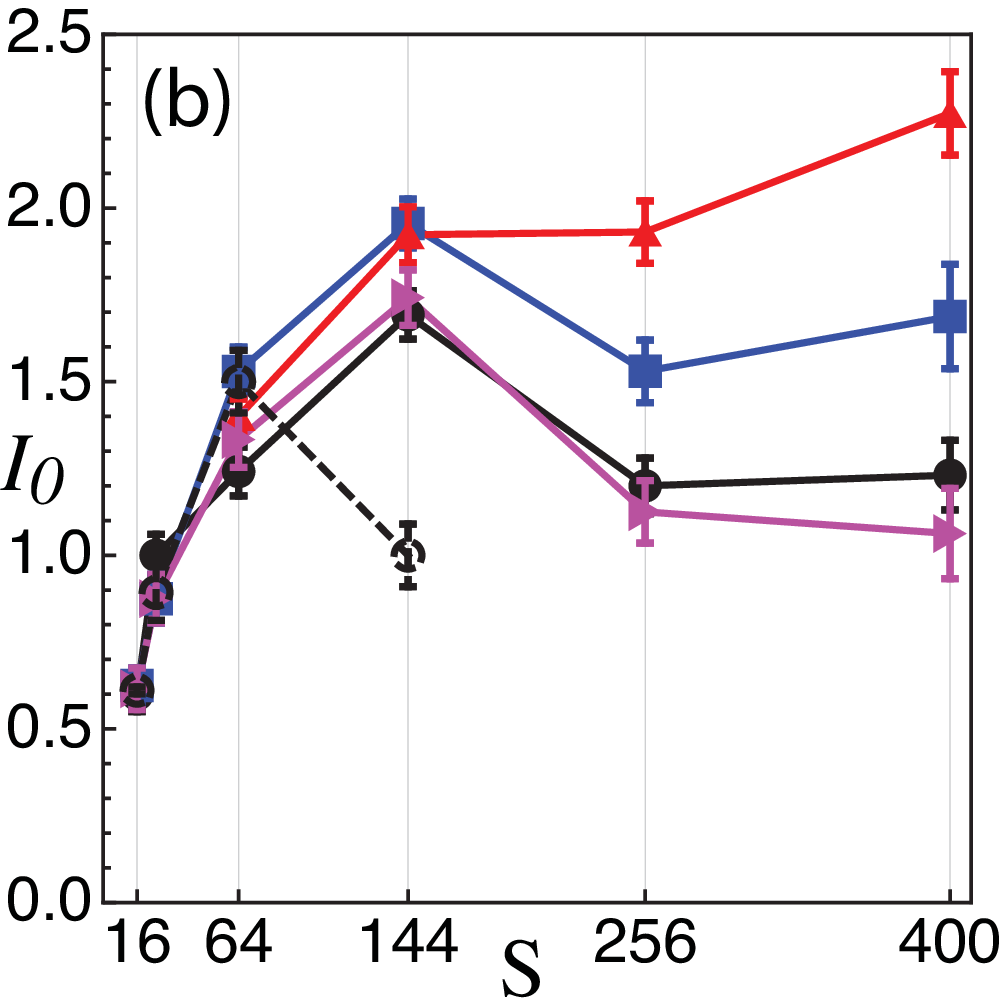}
  \includegraphics[width=0.32\columnwidth]{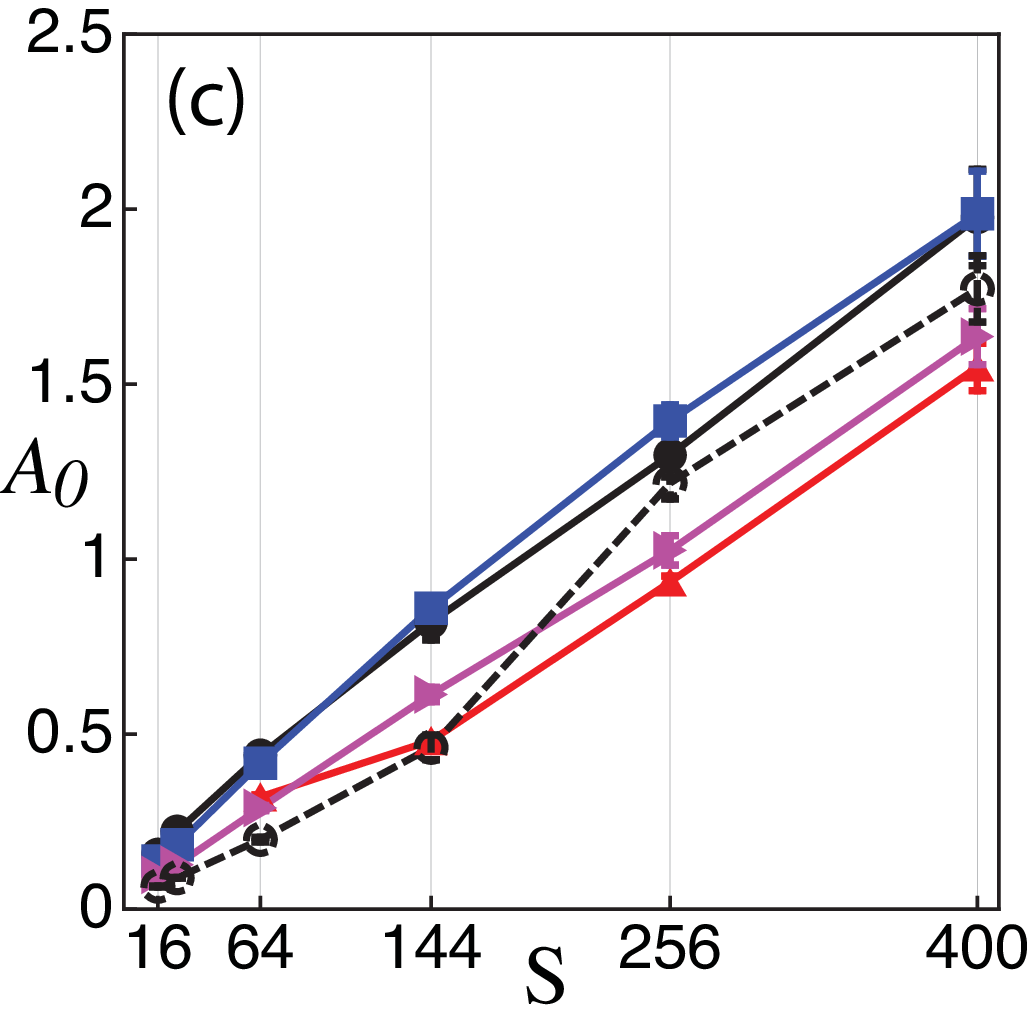}
 \end{center}
 \caption{
 		 		(a)
 			LFO frequencies, $\lfoo$, as a function of $S$, for different container lengths $l_x$, and shaking amplitudes $\ina$, as given in the inset.
		(b)
 			Intensity of $\lfoo$, $I_0$, defined as the height from the assymptotic low-frequencies value of the $z_{cm}(t)$ spectra to the broad peak, for the same data as (a).
		(c)
 			Amplitude of the LFOs, defined as the standard deviation of $z_{cm}(t)$, as a function of $S$, for the same data as (a).
 		}
 \label{figLFOs}
\end{figure}

All spectra present two main features: the expected delta-like peak at $\ino$ and its harmonics, and a broad peak one to two orders of magnitude lower, corresponding to the LFOs. The LFO frequency, $\lfoo$, is defined as the frequency of the maximum of this broad peak. After observing the different spectra it becomes evident that $\lfoo$ depends on the energy injection parameters. Figure \ref{figLFOs}a shows $\lfoo(S)$ for different $l_x$ and $\ina$, remarkably scaling all LFO data. Notice that $\lfoo$ decreases as $S$ increases, i.e., the collective grain movement becomes slower as the shaking gets faster. The decay is faster than inverse linear, and can be fitted by a $-\tfrac{3}{2}$ power with a $5\%$ error (not shown). Let us also notice that the length of the container makes no discernible difference, as long as the system stays in the Leidenfrost state; the decreased data in the $l_x = 20$ case are due to the Leidenfrost-convection transition. The collapse of the different amplitude curves is very good for $\ina = 0.4$ and $\ina = 1.0$, while for $\ina = 4.0$ data slightly deviates. We interpret this decrease as the influence of the undulations state in the Leidenfrost regime; notice that for $S \sim 64$ and $\ina = 4.0$ the system is almost at the boundary between both states (see Figure \ref{figPhase1}a).


In order to quantify the relevance of the LFOs, we define the relative intensity of the $\omega_0$ peak, $I_0$, as the normalised distance from the low frequencies asymptotic limit to the maximum of the broad peak. Figure \ref{figLFOs}b shows $I_0(S)$ for different $l_x$ and $\ina$. Although the dependency is not straightforward, it can be seen that LFOs become increasingly distinguishable from other movements until $S \sim 144$, after which there is a decline, except for the highest amplitude case. Already at $S = 25$ oscillations should be discernible in the spectra as a peak twice as big as the low-frequency asymptotic limit. $\ina$ is seen to have a pronounced effect on $I_0$; higher amplitudes of oscillation lead to more pronounced LFO peaks.

Finally, we define the amplitude of the LFOs, $\lfoa$, as the standard deviation of $z_{cm}(t)$: $\lfoa \equiv \sigma(z_{cm}(t))$. Data is considered only after $t = 1000$, to disregard transient states. Figure \ref{figLFOs}c shows $\lfoa(S)$ increasing in an almost linear way. The curves coincide, within their error, for $\ina = 0.4$ and $\ina = 1.0$, while for all other cases $\lfoa$ is consistently smaller. Nevertheless, $S$ makes all curves comparable, further confirming its relevance for this system.

\subsection{LFO's in convective state}

\begin{figure}
 \begin{center}
 \includegraphics[width=0.98\columnwidth]{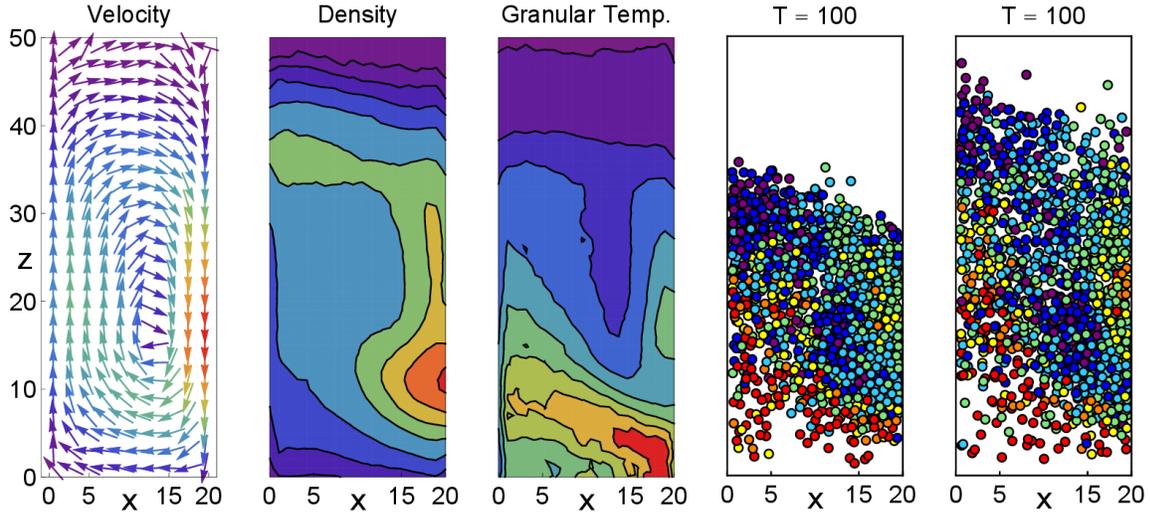}
 \end{center}
 \caption{
 		(a)
 			Averaged velocity field of an $l_x = 20$ system in the convective state, for $\ina = 1.0$ and $S = 144$ ($\ino = 12$). The colour of the arrows corresponds to the average speed, increasing from blue, green, yellow, until red.
 		(b)
 			Average density field of the system in (a). Colour scale from blue (low densities) to red (high densities).
 		(c)
 			Average granular temperature field, as defined in main text.
 		(d, e)
 			Two snapshots of the system taken at the minimum (d) and maximum (e) of a low-frequency oscillation. Colour corresponds to the particles kinetic energy.
 		}
 \label{figConvection}
\end{figure}

We now consider in detail the peculiar change of behaviour of $\lfoo(S)$ and $\lfoa(S)$ for $S \sim 144$ in the $l_x = 20$ case. This is a sign of the Leidenfrost-convection transition, still present at this container length (see Figure~\ref{figPhase1}a). During convection, $z_{cm}$ becomes a less relevant quantity, as there is no longer horizontal homogeneity. Nevertheless it is still possible to identify LFOs, even if the oscillations are entangled with the convective flow. The presence of LFOs in the convective regime should not be surprising if one notices that it also presents the essential feature of the Leidenfrost state: a high density, low temperature region suspended over a low density, highly agitated one, although there is an additional low density, highly convective zone above. Our model, derived in Section~3 below, suggests that when density inversion is present, LFOs exist. Figure~\ref{figConvection} presents several different fields and snapshots that show that, indeed, density inversion is present in the convective regime, in addition to the horizontal inhomogeneity. All data is taken from the same simulation, and fields are time-averaged over $100T$ after an initial transient of $1000T$, with data taken every $0.05T$. The average velocity field, Figure~\ref{figConvection}a, clearly shows the presence of convective flow, with a small downwards band and a wider upwards region. Particles agglomerate at the bottom of the downwards flux side, as can be seen from the average density field (Figure ~\ref{figConvection}b), and the two snapshots (Figures ~\ref{figConvection}d and \ref{figConvection}e). This happens when downwards and upwards particles collide, leading to a high granular temperature region (Figure ~\ref{figConvection}c). Note, then, that both sides correspond to low density, high temperature regions sustaining high density, lower temperature ones, although the density and temperature profiles vary considerably from left to right. The profile is more similar to the Leidenfrost case in  the upwards 
flow region (left in the shown figures), as in the downwards flow region the high density area presents a comparable, although lower temperature to the low density region below.

\subsection{Summary}

Having possible experimental realisations in mind, the general picture is that LFOs are easier to observe for higher amplitude and frequencies of oscillation of the box, while keeping $l_x = l_y$ small; it is at these configurations that LFOs have the highest amplitudes and better defined frequencies, as quantified by $\lfoa$ and $I_0$, respectively. Let us now remember that at this limit we also observed the most clear phase separation in the Leidenfrost state, with distinct low and high density regions. In our model, presented next, the separation of the phases and the confinement of the system to a one-dimensional geometry implies the existence of LFOs, and the frequency is essentially determined by the ratio of the low and high densities.

\section{Continuum model}
\label{sec:model1}

After observing the collective movement of the particles in the column geometry, an oscillator-like description naturally comes to mind. The two coexisting frequencies observed in the spectra suggest a forced oscillator model, with clearly defined forcing and response frequencies. In the following we derive such frequency behaviour from a continuum description of the granular media. We begin by considering Cauchy's equations for mass and momentum conservation:
\ifscaling
\else
\begin{align}
	\label{eq:mass_balance}
	&D_{t} \rho + \rho (\nabla \cdot \vec{u}) = 0, \\
	\label{eq:momentum_balance}
	&D_{t} (\rho\vec{u}) = \nabla\cdot \hat \sigma - \rho g \hat z,
\end{align}
where $\rho$ corresponds to the material density, $\vec{u} = \{u,v,w\}$ is the velocity vector, $\hat \sigma$ the stress tensor and $g$ the gravitational acceleration in the downwards direction, $-\hat z$. Furthermore, the material derivative is defined as $D_{t} \equiv \partial_{t} + \vec{u} \cdot \nabla$. We consider the same scaling as in simulations, with length scales in units of particle diameters $\tilde d$, time units given by gravity $\tilde t_g = (\tilde d / \tilde g)^{1/2}$, as also $\tilde \rho_p$, taken as the mass density of a single particle, $\tilde \rho_p = \tilde m_p /\tilde V_p$, with $\tilde V_p = \tfrac{1}{6}\pi\tilde d^3$.

\fi

\ifscaling
\else

As has been observed in simulations, the dynamics of the system in the column limit is effectively one-dimensional. This immediately suggest the consideration of $\rho = \rho(z,t)$, $\vec u = w (z,t)\hat z$ and $\hat \sigma = \sigma_{zz}(z,t)$. Substituting in \eqref{eq:mass_balance} yields
\begin{equation}
	\label{eq:mass_balance_1d}
	\partial_t \rho + w \partial_z \rho + \rho \partial_z w = 0.
\end{equation}
Furthermore, expanding \eqref{eq:momentum_balance}, and using \eqref{eq:mass_balance_1d}, one reaches a one-dimensional momentum conservation equation
\begin{equation}
	\label{eq:momentum_1D}
	\rho \partial_t w = \partial_z \sigma_{zz} - \rho g.
\end{equation}

\fi

\subsection{Two phases approximation}

In order to solve \eqref{eq:momentum_1D} it would be necessary to know both the density and the velocity profiles, $\rho(z,t)$ and $w(z,t)$. Our approach consists in eliminating the $z$-dependence from \eqref{eq:momentum_1D} by integrating in the vertical direction, and taking a first order approximation of the density profile $\rho(z,t)$, and average values for the vertical velocity profile $w(z,t)$. We begin by integrating \eqref{eq:momentum_1D} in the vertical direction
\begin{align}
	\label{eq:integrated_momentum}
	\int_{b(t)}^{s(t)} \rho \partial_t w\,\text{d}z &= \int_{b(t)}^{s(t)} \partial_z \sigma_{zz}\,\text{d}z - g \int_{b(t)}^{s(t)} \rho\,\text{d}z,
\end{align}
with the bottom boundary, $b(t)$, and top boundary, $s(t)$, dependent on time, due to the movement of the bottom wall and the free surface at the top.

\begin{figure}
	\begin{center}
      \includegraphics[width=0.98\columnwidth]{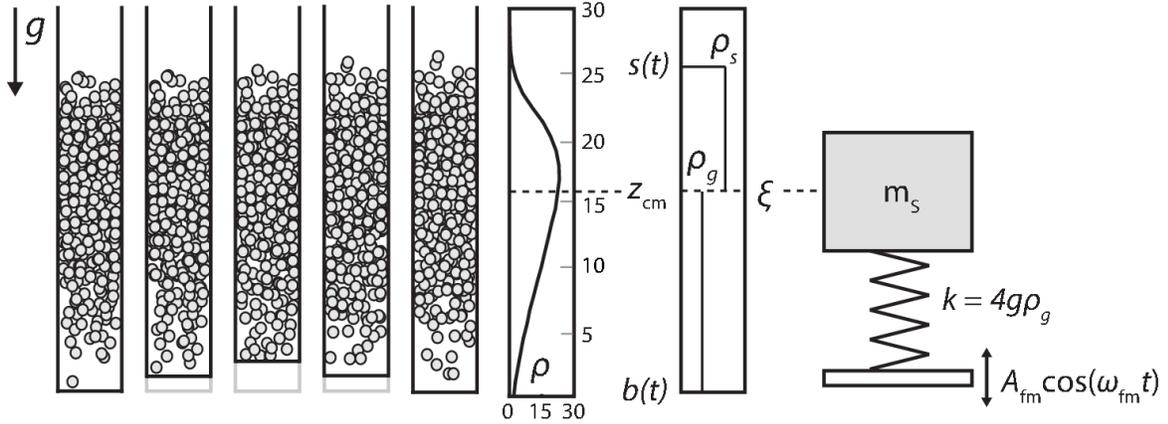}
 	\end{center}
 	\caption{
 	From left to right, snapshots from simulations showing an LFO period, at phases $0$, $\pi/2$, $\pi$, $3\pi/2$ and $2\pi$; the corresponding time averaged density profile, a representation of the two phases approximation made for the continuum equations, and finally a schematic representation of the model. The dashed line shows the position of the centre of mass, $z_{cm}$, which in the model corresponds to the position of the interface between the two phases, $\xi$, which also corresponds to the position of the mass of a forced harmonic oscillator.
 	 }
 \label{figModel02}
\end{figure}

The approximation of $\rho(z)$ consists in dividing the system in two separate, constant density regions, inspired by the measured Leidenfrost state density profile. Let us remember that this approximation becomes increasingly better as $S$ increases and $\ina$ decreases, as shown in Figure~\ref{figPhase1}b. Consequently, a low density region is defined where $\rho(z,t) = \rho_g(t)$, for $z < \xi$; and a high density region where $\rho(z,t) = \rho_s(t)$, for $z > \xi$, with $\xi = \xi(t)$ the position of the interface between the two regions. Figure~\ref{figModel02} shows a schematic representation of this approximation, and the origin of its motivation. It then follows that the first integral in \eqref{eq:integrated_momentum} can be expanded as
\begin{equation}
	\label{eq:divided_integral_1}
	\int_{b(t)}^{s(t)} \rho \partial_t w\,\text{d}z = \rho_g \int_{b(t)}^{\xi(t)} \partial_t w\,\text{d}z + \rho_s \int_{\xi(t)}^{s(t)} \partial_t w\,\text{d}z.
\end{equation}
Analogously, the third integral in \eqref{eq:integrated_momentum} becomes
\begin{equation}
	\label{eq:divided_integral_2}
	g \int_{b(t)}^{s(t)} \rho \text{d}z = g \rho_g \int_{b(t)}^{\xi(t)} \text{d}z + g \rho_s \int_{\xi(t)}^{s(t)} \text{d}z = g \rho_g h_g + g \rho_s h_s,
\end{equation}
with $h_g(t) \equiv \xi(t) - b(t)$ the height of the gaseous region, and $h_s(t) \equiv s(t) - \xi(t)$ the height of the solid region. 

Notice that the second integral in \eqref{eq:integrated_momentum}, corresponding to the stress term, is a perfect integral, and thus only the stress boundary conditions are needed for its evaluation. We assume the stress through the system to be continuous in $z$, and thus it is not necessary to evaluate $\sigma_{zz}$ at the interface position $\xi(t)$. Thus, from \eqref{eq:integrated_momentum} we finally obtain:
\begin{equation}
	\label{eq:integrated_momentum_2}
\rho_g \int_{b(t)}^{\xi(t)} \partial_t w\,\text{d}z + \rho_s \int_{\xi(t)}^{s(t)} \partial_t w\,\text{d}z = \sigma_{zz}(z=s) - \sigma_{zz}(z=b) - g \rho_g h_g - g \rho_s h_s.
\end{equation}

\subsection{Boundary conditions}

It now becomes necessary to specify the boundary conditions. The shaking of the container implies that $b(t) = \inam \sin(\inom t)$, with $ \inam$ and $\inom$ the amplitude and frequency of energy injection in the model. At the top, s(t), we consider a free surface, and thus the kinematic boundary conditions are given by
\begin{align}
	\label{eq:kinetic_boundary_1}
	& w(b(t),t) = v_b = \inam \inom \cos(\inom t) \\
	\label{eq:kinetic_boundary_2}
	& w(s(t),t) = \partial_t s
\end{align}

Furthermore, the stress at the bottom and top of the granular media are needed. The free surface at the top is straightforward: $\sigma_{zz}(z=s) = 0$. At the bottom, on the other hand, we divide the stress contribution in two: mean ($\sigma_b^0$) and fluctuating ($\sigma_b$) terms, where the mean term is straightforward: $\sigma_b^0 = Mg/\eta$, with $M$ the total mass of the system, $M = Nm$; and $\eta$ the area of the base of the container, $\eta = l_xl_y$.

For the fluctuating part of the stress, $\sigma_b$, we first consider the force applied to the granular medium by the moving bottom:
\begin{equation}
  \label{newton}
	\sigma_b = \frac{F_b}{\eta} = d_t(m_bv_b)
\end{equation}
with $m_b$ the mass being pushed by the bottom wall. In order to obtain $m_b$, let us consider a moving platform of surface area $\eta$ pushing an ideal, incompressible gas of density $\rho_g$, in analogy to the moving box and the low density region observed in our system. Accordingly, the mass pushed by the box in time is given by $dm_p = \rho_g \eta v_b dt$. Notice that this is valid for high $S$, where gravity effects on the dynamics of the particles can be ignored. Integrating, we directly get that $m_p = \rho_g \eta \inam \sin(\inom t)$. Substituting in \eqref{newton}:
\begin{equation}
	F_b = d_t(m_pv_b) = \rho_g \eta \inam^2 \inom^2 (-\sin^2(\inom t) + \cos^2(\inom t)) = \rho_g \eta \inam^2 \inom^2 \cos(2\inom t).
\end{equation}
Notice that we have naturally obtained $\inam^2 \inom^2 \equiv \Sm g$ as the amplitude of the force applied by the oscillating bottom, further suggesting that the shaking strength is the relevant parameter for the system in the high $S$ limit. It then follows, from \eqref{newton}, that:
\begin{equation}
	\label{eq:boundary_bottom}
	\sigma_b = g \rho_g \Sm \cos(2\inom t).
\end{equation}
Finally, substituting the stress boundary values in (\ref{eq:integrated_momentum_2}), we obtain:
\begin{equation}
	\label{integrated_momentum_3}
\rho_g \int_{b(t)}^{\xi(t)} \partial_t w\,\text{d}z 
+ \rho_s \int_{\xi(t)}^{s(t)} \partial_t w\,\text{d}z = - g \rho_g \Sm \cos(2\inom t) - g \frac{M}{\eta} - g \rho_g h_g - g \rho_s h_s.
\end{equation}

\subsection{Height averaging}

The remaining two integrals in \eqref{integrated_momentum_3} involve the velocity profile, $w = w(z,t)$, which varies in the vertical direction. In order to solve these integrals we height-average, that is, for a given quantity $f(z)$, we consider its average value
\begin{equation}
	\label{eq:depth_averaging}
	\bar f \equiv \frac{1}{h} \int_{b(t)}^{s(t)} f\,\text{d}z = \frac{1}{h_g} \int_{b(t)}^{\xi(t)} f\,\text{d}z + \frac{1}{h_s} \int_{\xi(t)}^{s(t)}f\,\text{d}z.
\end{equation}
Notice that, from the first integral in \eqref{integrated_momentum_3}, $f$ would correspond to $\partial_t w$. Thus, before applying \eqref{eq:depth_averaging}, we express the integral as a total time derivative. Considering that the boundaries are time dependent, it becomes necessary to use Leibniz integration rule, and thus the first integral in \eqref{integrated_momentum_3} can be expressed as
\begin{equation}
  \label{eq:integral_1}
  \int_{b(t)}^{\xi(t)} \partial_t w\,dz = g S_m \cos^2(\inom t)  - w(z=\xi) d_t \xi  + d_t \int_{b(t)}^{\xi(t)} w\,dz
\end{equation}
Analogously, the second integral in \eqref{integrated_momentum_3} becomes, after using \eqref{eq:kinetic_boundary_2},
\begin{equation}
  \label{eq:integral_2}
  \int_{\xi(t)}^{s(t)} \partial_t w\,dz = w(z=\xi) d_t \xi - (d_t s)^2 + d_t \int_{\xi(t)}^{s(t)} w\,dz,
\end{equation}
Substituting \eqref{eq:integral_1} and \eqref{eq:integral_2} in \eqref{integrated_momentum_3}, and using \eqref{eq:depth_averaging}, we finally obtain:
\begin{align}
  \label{all_1}
	- \rho_g w(z=\xi) d_t \xi
      + \rho_g d_t (h_g \bar w_{g}) 
      + \rho_s w(z=\xi) d_t \xi
      - \rho_s (d_t s)^2
      + \rho_s d_t (h_s \bar w_{s})
      =\notag\\ 
      - \tfrac{1}{2} g \rho_g \Sm ( 3 \cos(2 \inom t) + 1 )
      - g \frac{M}{\eta}
      - g \rho_g h_g
      - g \rho_s h_s
\end{align}

Based on the behaviour observed in simulations, we now assume that the high density region is incompressible. This implies that $d_t h_s = 0$, as also that the velocity of the continuum media at the interface position is equivalent to the velocity of the interface, and hence to the velocity of the surface, that is, $w(z=\xi) = \bar w_s = d_t s = d_t \xi$. Thus \eqref{all_1} becomes
\begin{equation}
	- \rho_g (d_t \xi)^2
      + \rho_g d_t (h_g \bar w_g) 
      + \rho_s h_s d_{tt} \xi
      = 
      - \tfrac{1}{2} g \rho_g \Sm \left( 3 \cos(2 \inom t) + 1 \right)
      - g \frac{M}{\eta}
      - g \rho_g \xi
      - g \rho_s h_s.
\end{equation}
Furthermore, we now use the fact that $h_g(t) = \xi(t) - b(t)$. Thus, substituting and dividing by $\rho_s$, we obtain
\begin{align}
\label{eq:full_before_approx}
       - \frac{\rho_g}{\rho_s} (d_t \xi)^2
	+ \frac{\rho_g}{\rho_s} d_t (\xi \bar w_g) 
	- \frac{\rho_g}{\rho_s} d_t (\inam \sin(\inom t) \bar w_g)
      + h_s d_{tt} \xi
      = \notag \\
      \frac{g \rho_g}{2\rho_s} \Sm ( 3 \cos(2 \inom t) + 1 )
      + \frac{g M}{\eta \rho_s}
      - \frac{g \rho_g}{\rho_s} \xi
      - g h_s
\end{align}

\subsection{First order approximations}

It now becomes relevant to consider the relative importance of each of these terms, in the region of phase space where simulations show that LFOs are present, that is, for $S \gg 1$. First, we consider that $\rho_g/\rho_s \sim \mathcal{O}(\epsilon)$, a condition that holds better for $S \gg 1$ and low $A_f$, as shown in Figure~\ref{figPhase1}b. On the other hand, $\xi \sim 10 \sim \mathcal{O}(1/\epsilon)$, as can be seen from Figure \ref{figZCM}. Furthermore, we measure from simulations that $\delta_t\xi\sim0.2\sim\mathcal{O}(\epsilon)$ and $\delta_{tt}\xi\sim0.1\sim\mathcal{O}(\epsilon)$, meaning that the dynamics of the LFOs are considerably lower than the typical velocity of grain diameters per gravity timescale, as can also be deduced by the previously obtained frequencies $\omega_0$. Let us also notice that $h_s \sim h_g \sim 8 \sim \mathcal{O}(1/\epsilon)$, again, from Figure~\ref{figPhase1}b. Finally, from simulations we obtain that $\bar w_g \sim 0.2 \sim \mathcal{O}(\epsilon)$, and $d_t \bar w_g \sim 0.02 \sim \mathcal{O}(\epsilon^2)$. Taking into account all these considerations, it becomes straightforward to see that the first term in \eqref{eq:full_before_approx} is $\mathcal{O}(\epsilon^3)$, the second term is at most $\mathcal{O}(\epsilon^2)$, the third is then $\mathcal{O}(\epsilon)$, and the fourth term is $\mathcal{O}(1)$. Moreover, all terms on the right side are $\mathcal{O}(1)$. Thus, disregarding small terms in \eqref{eq:full_before_approx}, after dividing by $h_s$, we obtain

\begin{equation}
\label{eq:osc}
	  d_{tt} \xi
      + \frac{g \rho_g}{m_s} \xi
      = 
      \frac{g \rho_g}{2m_s} \Sm ( 3 \cos(2 \inom t) + 1 )
      + g \left( m_g/m_s
      - 1\right) 
\end{equation}
where we have defined the mass of the solid region per unit base area $\eta$, $m_s \equiv h_s \rho_s$, and the equivalent of the gaseous region, $m_g = h_g \rho_g$. Equation \eqref{eq:osc} corresponds to a forced harmonic oscillator equation of the form:
\begin{equation}
\label{eq:osc2}
	d_{tt} \xi + \lfoom^2 \xi = F_0 \cos( 2 \inom t) + C,
\end{equation}
with natural frequency
\begin{equation}
  \label{omega}
  \lfoom^2 = \frac{g \rho_g}{m_s},
\end{equation}
amplitude of forcing $F_0 = \tfrac{3}{2} g \rho_g \Sm /m_s$, and constant $C = \tfrac{1}{2} \rho_g \Sm / m_s + g(m_g/m_s-1)$.

\subsection{Model and simulations comparison}

We have shown that, considering Cauchy's equations for continuum media, and making assumptions in concordance to the observed granular Leidenfrost state, the system becomes equivalent to a simple forced harmonic oscillator, expressed by \eqref{eq:osc2}. In this case, $\xi$ is the displacement of the centre of mass around the equilibrium position at $0$, $\lfoom$ the natural frequency of the system, and $F_0$ and $\inom$ the amplitude and frequency of the forcing. The analogy of the forcing with the granular column is straightforward: $\lfoom$ and $\lfoam$ would be equivalent to $\lfoo$ and $\lfoa$, respectively. Furthermore, we choose $\xi$ to correspond to $z_{cm}$, in order to directly compare with previous measurements.


Notice that the natural frequency $\lfoom$ does not explicitly depend on the forcing frequency $\inom$, as can be seen in \eqref{omega}. The implicit dependence comes from the variation of $\rho_g$ and $m_s$ with $\inom$, as observed in simulations, where, for fixed $S$, $\rho_g/\rho_s$ increases with $\ino$, giving the correct inverse proportionality of $\lfoom$ with $\inom$. Therefore, in order to obtain a frequency from the model, only $\rho_g$ and $m_s$ need to be specified, which we measure from simulations.

\begin{figure}
	\begin{center}
      \includegraphics[scale=0.5]{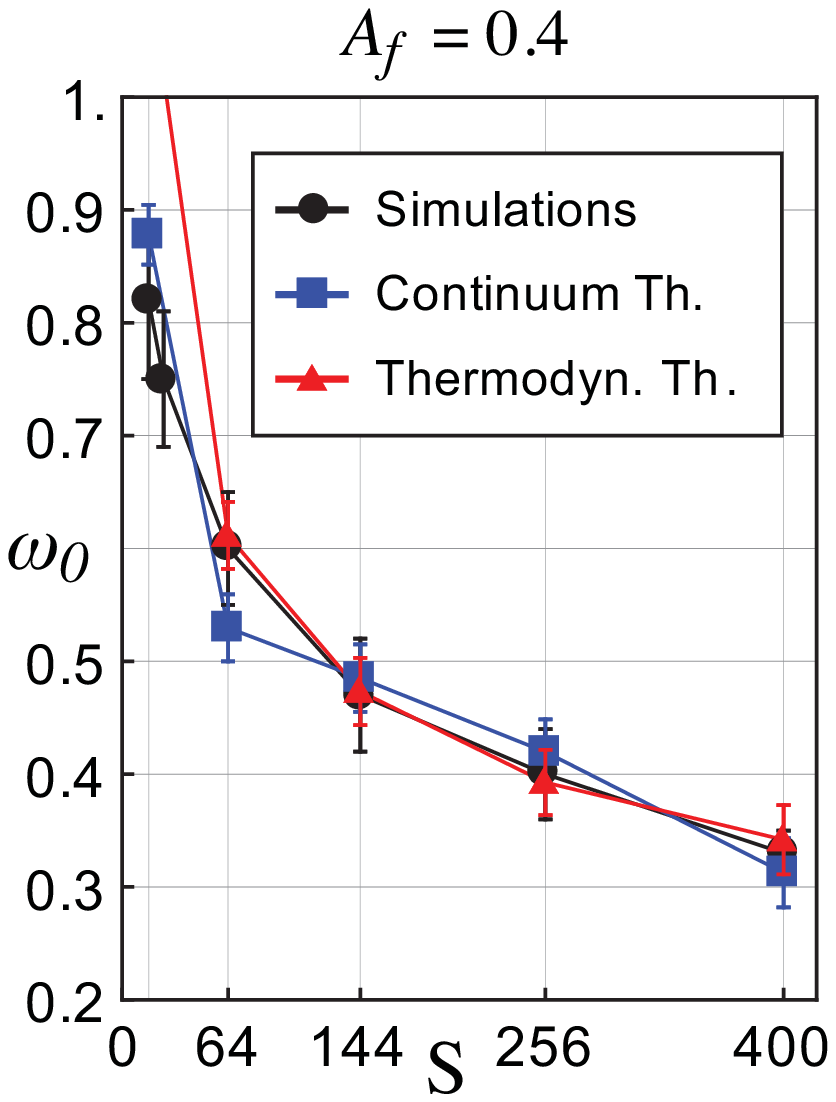}
      \includegraphics[scale=0.5]{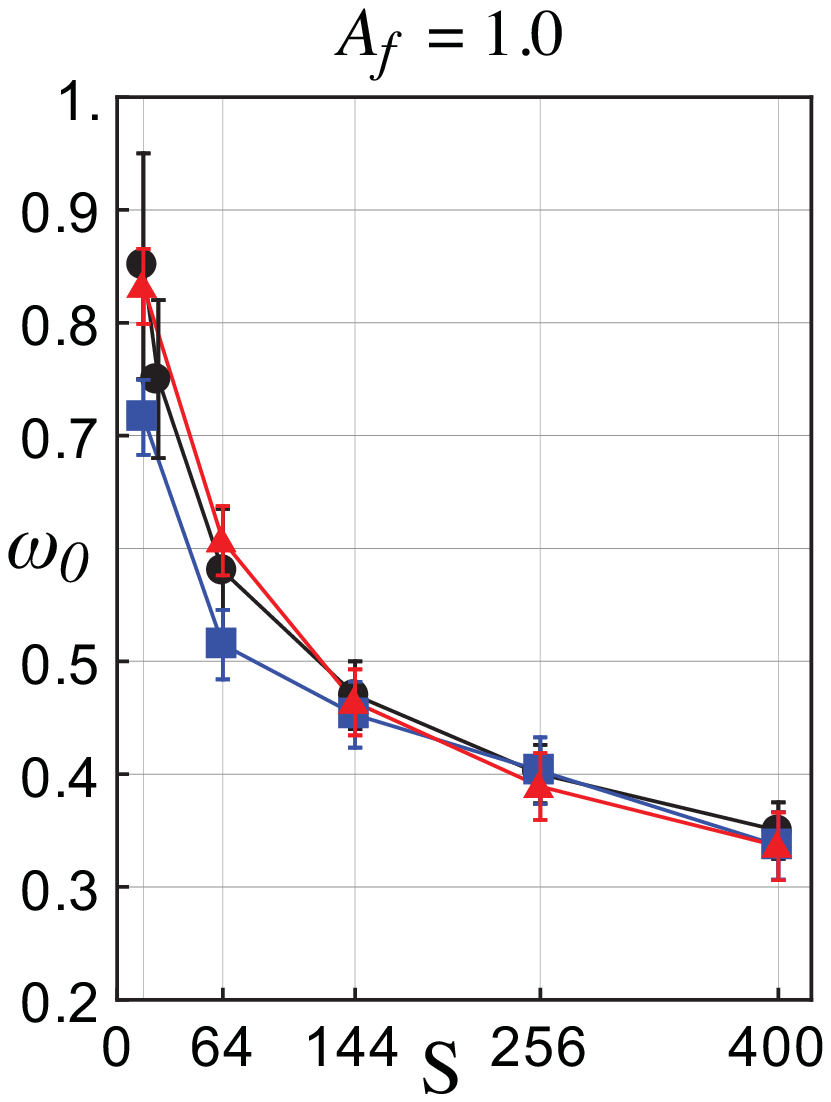}
      \includegraphics[scale=0.5]{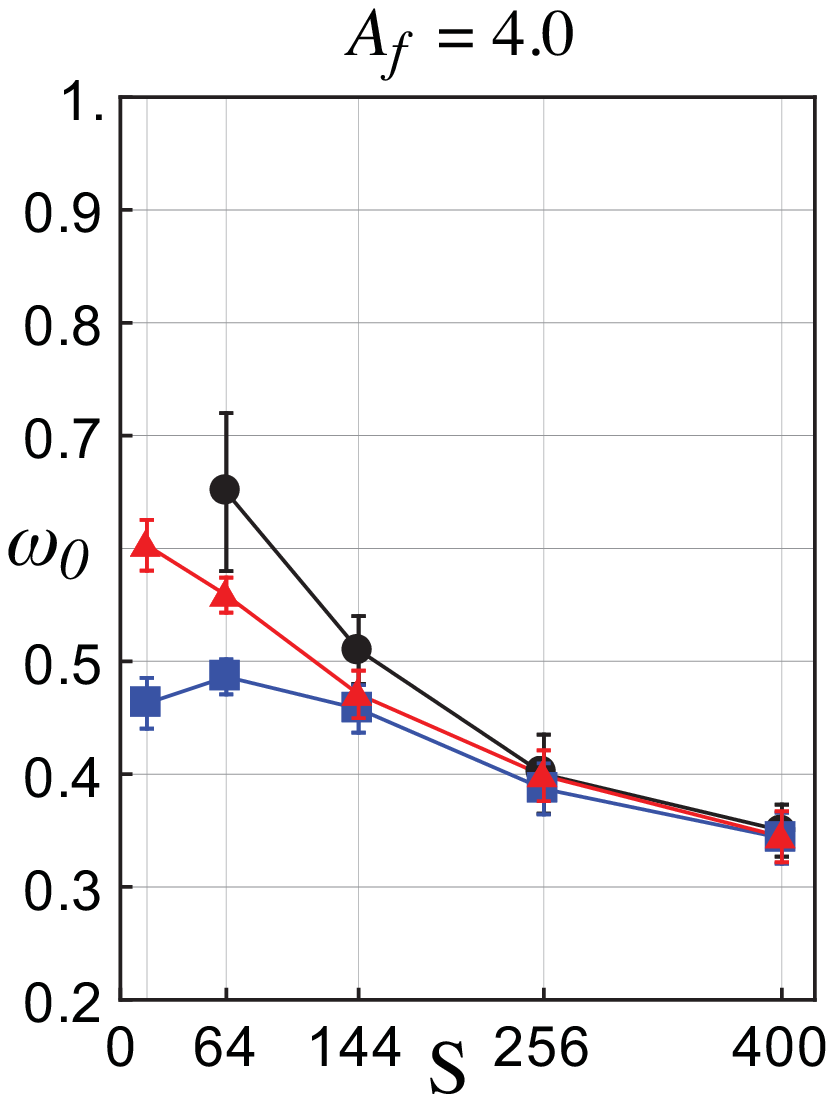}
 	\end{center}
 	\caption{
 	Low-frequency oscillation frequencies $\omega_0$, as a function of the dimensionless shaking strength $S$, for different box oscillation amplitudes: $A = 0.4$ (a), $A = 1.0$ (b) and $A = 4.0$ (c). All systems have $l_x = 5$, $F = 12$. Simulation (black) corresponds to frequencies obtained from fast-Fourier transform of the simulation data, while continuum and thermodynamic/kinetic theory data points (blue and red, respectively) are obtained from models presented in sections \ref{sec:model1} and \ref{sec:model2}, respectively, using data acquired from simulations.
 	}
 \label{figModel01}
\end{figure}

Both quantities can be obtained from $\rho(z)$, the density in the granular column as a function of height. In order to obtain an accurate average, we consider $\rho^* \equiv \rho(z-z_{cm}(t),t)$, which makes all profiles directly comparable. This is analogous, in the model, to centering the profiles at the interface between the two distinct regions. It is then straightforward to compute $\rho_g$ as the average value of the density for $z < z_{cm}$. On the other hand, $m_s$ we take as the total mass for $z > z_{cm}$, taking care not to count particles that are in free flight above the solid region, as they do not have influence on the oscillator dynamics. This implies that although the center of our profiles is $z_{cm}$, $m_g$ is not equal to $m_s$.

The comparison between the frequencies obtained in simulations and from the model is presented in Figures \ref{figModel01}a-c. For low amplitudes, $\ina \le 1.0$, and high frequencies, $S \ge 144$, the agreement between the frequencies is within the error bars. For lower $S$, or higher amplitudes, the assumption of two distinct phases, as also the approximation of $\xi \sim 1/\epsilon$, become less justified, resulting in the model consistently underpredicting the frequencies, with more than $50\%$ disagreement at the point of the bouncing bed-Leidenfrost transition for $A = 4.0$. We believe that the prediction could be improved by considering more complex density profiles, as also by including terms of lower orders, although this exceeds the scope of our work. In general terms, the resulting one-dimensional model turns out to be a remarkable well approximation for high $\ino$ and low $\ina$, showing that this many-particle, out-of-equilibrium system actually behaves as a regular forced harmonic oscillator when confined in a column, in the corresponding energy injection region.

\section{Thermodynamic model}
\label{sec:model2}

Remarkably, it is possible to obtain another accurate expression for $\omega_0$ using a completely different approach, considering basic concepts from thermodynamics. Assuming a spring-like behaviour, the natural frequency of our medium is given by $\omega_0^2 = k/m_s$, with $k$ the stiffness constant of the spring-like medium, and $m_s$ the mass sustained by the spring. We also know that $k = \eta B/h_g$, with $B$ the bulk modulus, $\eta$ the area of the spring, and $h_g$ it's height at rest. Assuming an adiabatic ideal gas, it is possible to relate the bulk modulus with the pressure, $B = \gamma P_0$, with $\gamma$ the adiabatic index. Notice that the ideal gas approximation is being used only for the gaseous region of the Leidenfrost regime, where densities are low and no significant correlation of the particles is observed. We then obtain:
\[
	\omega_0^2 = \frac{\gamma \eta P_0}{h_g m_s}.
\]

The pressure $P_0$ is taken as the force caused by the solid mass the spring sustains, $m_s$, divided by the area of the container: $P_0 = m_s g / \eta$. Thus, we finally reach:
\begin{equation}
\label{omega2}
	\omega_0^2 = \frac{\gamma g}{h_g}.
\end{equation}

This significantly simple expression is remarkably accurate when compared with simulation measurements. Figures \ref{figModel01}a-c also show $\omega_0(S)$ for this model, taking $h_g$ to be the same as in the previous section, $z_{cm}$. The adiabatic index is considered as $
\gamma = 1.67$, the theoretical value for an ideal monoatomic gas. The agreement is again within error bars for high frequencies, and deviates considerably for lower frequencies, except for the $A = 1.0$ case, where low frequencies are also captured. Relating the two obtained LFOs frequencies, equations (\ref{omega}) and (\ref{omega2}), one obtains $\gamma = m_g/m_s$; the interpretation of this result remains a challenge.

\section{Conclusions}

A vertically vibrated bed of grains presents low-frequency oscillations (LFOs) due to the decoupling of the driving frequency and the dynamics of a high density region suspended by a lower density one. The relevance of these oscillations increases as the distinction between the two densities increases, that is, proportional to the frequency and inversely to the amplitude of oscillation of the system container. The LFO frequencies are inversely proportional to the driving frequency, and follow a common power law for a range of amplitudes. The amplitude of the oscillations, on the other hand, increases in an almost linear way with the frequency.

Event-driven simulations give an overall excellent qualitative and quantitative agreement with experiments and soft-particle simulations done in wider systems, although they show discrepancies in some critical transition values. We remark that the hard-sphere approximation can be meaningful even in systems with very high-density regions, as present in the Leidenfrost state. The considerable speed advantage makes it extremely useful, and sometimes the only means to systematically study high dimensional parameter spaces.

Starting from Cauchy's equations for conservation of mass and momentum, integrating in the vertical direction and assuming two distinct low and high constant density regions, it is possible to reproduce the frequency behaviour observed in simulations. That is, a forced harmonic oscillator, with the natural frequency proportional to the ratio of the densities. This simple model is able to predict the natural LFO frequency for high excitation frequencies, where in fact the two phases are well separated. The non-linear terms, discarded in our analysis, should provide the necessary corrections for lower frequencies, as well as the consideration of a more realistic density profile. A second approach, using thermodynamic arguments, also gives a remarkably accurate expression for the frequencies, although in this case just a simple mass-spring system behaviour was assumed. The quantitative agreement of both models is nevertheless remarkable, taking into account the low number of particles involved, and the presence of very high and low density regions.

Further insight could be gained by appropriately coarse graining the granular medium in order to obtain stress fields, which would directly relate both models. A point of interest, not studied here, is how well do kinetic theory predictions hold in such a system, taking into account the reduced container size, the small number of particles and the presence of considerably different densities. Current work is being done on verifying the consistency of macroscopic fields obtained by theoretical arguments and coarse-grained simulational data.

We suggest that LFOs, here shown to be ubiquitous to vertically vibrated density inverted systems, could play a fundamental role in the Leidenfrost-convection transition. More specifically, LFOs could be the primary source of density fluctuations observed before convection is triggered, when one region of the system oscillates at a different phase than another. Understanding this will need further simulation and experimental research.

\section{Acknowledgements}

The authors wish to acknowledge Thomas Weinhart for a careful revision and important suggestions regarding Section 3. This work was financially supported by the NWO-STW VICI grant 10828.

\section*{References}

\bibliographystyle{unsrt}
\bibliography{libraryLFOs}

\begin{thebibliography}{10}

\bibitem{olafsen_clustering_1998}
J.~S. Olafsen and J.~S. Urbach.
\newblock Clustering, order, and collapse in a driven granular monolayer.
\newblock {\em Physical Review Letters}, 81(20):4369, 1998.

\bibitem{melby_dynamics_2005}
P.~Melby, F.~Vega Reyes, A.~Prevost, R.~Robertson, P.~Kumar, D.~A. Egolf, and
  J.~S. Urbach.
\newblock The dynamics of thin vibrated granular layers.
\newblock {\em Journal of Physics: Condensed Matter}, 17(24):S2689--S2704,
  2005.

\bibitem{melo_transition_1994}
F.~Melo, P.~Umbanhowar, and H.~L. Swinney.
\newblock Transition to parametric wave patterns in a vertically oscillated
  granular layer.
\newblock {\em Physical Review Letters}, 72(1):172, 1994.

\bibitem{luding_simulations_1996}
S.~Luding, E.~Clément, J.~Rajchenbach, and J.~Duran.
\newblock Simulations of pattern formation in vibrated granular media.
\newblock {\em Europhysics Letters ({EPL)}}, 36(4):247--252, 1996.

\bibitem{tennakoon_vertical_1998}
S.~G.~K. Tennakoon and R.~P. Behringer.
\newblock Vertical and horizontal vibration of granular materials: Coulomb
  friction and a novel switching state.
\newblock {\em Physical Review Letters}, 81(4):794, 1998.

\bibitem{medved_connections_2002}
M.~Medved.
\newblock Connections between response modes in a horizontally driven granular
  material.
\newblock {\em Physical Review E}, 65(2):021305, 2002.

\bibitem{ahmad_observation_1973}
K.~Ahmad and I.~J. Smalley.
\newblock Observation of particle segregation in vibrated granular systems.
\newblock {\em Powder Technology}, 8(1–2):69--75, 1973.

\bibitem{kudrolli_size_2004}
A.~Kudrolli.
\newblock Size separation in vibrated granular matter.
\newblock {\em Reports on Progress in Physics}, 67(3):209--247, 2004.

\bibitem{luding_cluster-growth_1999}
S.~Luding and H.~J. Herrmann.
\newblock Cluster-growth in freely cooling granular media.
\newblock {\em Chaos: An Interdisciplinary Journal of Nonlinear Science},
  9(3):673, 1999.

\bibitem{rivas_sudden_2011}
N.~Rivas, S.~Ponce, B.~Gallet, D.~Risso, R.~Soto, P.~Cordero, and N.~Mujica.
\newblock Sudden chain energy transfer events in vibrated granular media.
\newblock {\em Physical Review Letters}, 106(8):088001, 2011.

\bibitem{thomas_identifying_1989}
B.~Thomas, M.~O. Mason, Y.~A. Liu, and A.~M. Squires.
\newblock Identifying states in shallow vibrated beds.
\newblock {\em Powder Technology}, 57(4):267--280, 1989.

\bibitem{douady_subharmonic_1989}
S.~Douady, S.~Fauve, and C.~Laroche.
\newblock Subharmonic instabilities and defects in a granular layer under
  vertical vibrations.
\newblock {\em Europhysics Letters ({EPL)}}, 8(7):621--627, 1989.

\bibitem{wassgren_vertical_1996}
C.~R Wassgren, C.~E Brennen, and M.~L Hunt.
\newblock Vertical vibration of a deep bed of granular material in a container.
\newblock {\em Journal of Applied Mechanics}, 1996.

\bibitem{meerson_close-packed_2003}
B.~Meerson, T.~Pöschel, and Y.~Bromberg.
\newblock Close-packed floating clusters: Granular hydrodynamics beyond the
  freezing point?
\newblock {\em Physical Review Letters}, 91(2):024301, 2003.

\bibitem{leidenfrost_aquae_1756}
J.~G. Leidenfrost.
\newblock De aquae communis nonnullis qualitatibus tractatus.
\newblock 1756.

\bibitem{eshuis_granular_2005}
P.~Eshuis, K.~van~der Weele, D.~van~der Meer, and D.~Lohse.
\newblock Granular leidenfrost effect: Experiment and theory of floating
  particle clusters.
\newblock {\em Physical Review Letters}, 95(25), 2005.

\bibitem{eshuis_phase_2007}
P.~Eshuis, K.~van~der Weele, D.~van~der Meer, R.~Bos, and D.~Lohse.
\newblock Phase diagram of vertically shaken granular matter.
\newblock {\em Physics of Fluids}, 19(12):123301, 2007.

\bibitem{lubachevsky_how_1991}
B.~D. Lubachevsky.
\newblock How to simulate billiards and similar systems.
\newblock {\em Journal of Computational Physics}, 94(2):255--283, 1991.

\bibitem{luding_granular_1995}
S.~Luding.
\newblock Granular materials under vibration: Simulations of rotating spheres.
\newblock {\em Physical Review E}, 52(4):4442--4457, 1995.

\bibitem{luding_how_1998}
Stefan Luding and Sean {McNamara}.
\newblock How to handle the inelastic collapse of a dissipative hard-sphere gas
  with the {TC} model.
\newblock {\em Granular Matter}, 1(3):113--128, 1998.

\bibitem{clerc_liquid-solid-like_2008}
M.~G. Clerc, P.~Cordero, J.~Dunstan, K.~Huff, N.~Mujica, D.~Risso, and
  G.~Varas.
\newblock Liquid-solid-like transition in quasi-one-dimensional driven granular
  media.
\newblock {\em Nat Phys}, 4(3):249--254, 2008.

\bibitem{mcnamara_energy_1998}
Sean {McNamara} and Stefan Luding.
\newblock Energy flows in vibrated granular media.
\newblock {\em Physical Review E}, 1998.

\bibitem{leidenfrost_fixation_1966}
J.~G. Leidenfrost.
\newblock On the fixation of water in diverse fire.
\newblock {\em International Journal of Heat and Mass Transfer},
  9(11):1153--1166, 1966.

\end{thebibliography}

\end{document}